\documentclass[%
twocolumn,
superscriptaddress,
nofootinbib,
 amsmath,amssymb,
 aps, prd
]{revtex4-1}

\usepackage{graphicx}
\usepackage{tensor}
\usepackage{siunitx}
\usepackage{xcolor} 
\usepackage{lipsum} 

\newcommand{\beq}{\begin{equation}}
\newcommand{\eeq}{\end{equation}}
\newcommand{\beqn}{\begin{eqnarray}}
\newcommand{\eeqn}{\end{eqnarray}}

\newcommand{\lo}{\mathrel{\raise.3ex\hbox{$<$}\mkern-14mu
    \lower0.6ex\hbox{$\sim$}}}
\newcommand{\go}{\mathrel{\raise.3ex\hbox{$>$}\mkern-14mu
    \lower0.6ex\hbox{$\sim$}}}

\newcommand{\WSU}{\affiliation{Department of Physics \& Astronomy,
	Washington State University, Pullman, Washington 99164, USA}}
\newcommand{\UNH}{\affiliation{Department of Physics, University of New Hampshire, 9 Library Way, Durham NH 03824, USA}}
\newcommand{\TAPIR}{\affiliation{TAPIR, Walter Burke Institute for Theoretical Physics, MC 350-17, California Institute of Technology, Pasadena, California 91125, USA}}
\newcommand{\Cornell}{\affiliation{Center for Radiophysics and Space Research, Cornell University, Ithaca, New York, 14853, USA}}
\newcommand{\MPI}{\affiliation{Max-Planck-Institut fur Gravitationsphysik, Albert-Einstein-Institut, D-14476 Potsdam, Germany}}

\begin{document}
\title{A comparison of momentum transport models for numerical relativity}
\author{Matthew D. Duez} \WSU
\author{Alexander Knight} \UNH
\author{Francois Foucart} \UNH
\author{Milad Haddadi} \WSU
\author{Jerred Jesse} \WSU
\author{Fran\c{c}ois H\'{e}bert} \TAPIR
\author{Lawrence E. Kidder} \Cornell
\author{Harald P. Pfeiffer} \MPI
\author{Mark A. Scheel} \TAPIR

\begin{abstract}
  The main problems of nonvacuum numerical relativity, compact binary mergers and stellar
  collapse, involve hydromagnetic instabilities and turbulent flows, so that kinetic energy
  at small scales lead to mean effects at large scale that drive the secular evolution.
  Notable among these effects is momentum transport.  We investigate two models of this
  transport effect, a relativistic Navier-Stokes system and a turbulent mean stress model,
  that are similar to all of the prescriptions that have been attempted to date for treating
  subgrid effects on binary neutron star mergers and their aftermath.  Our investigation
  involves both stability analysis and numerical experimentation on star and disk systems.
  We also begin the investigation of the effects of particle and heat transport on post-merger
  simulations.  We find that correct handling of turbulent heating is crucial for avoiding unphysical
  instabilities.  Given such appropriate handling, The evolution of a differentially rotating
  star and the accretion rate
  of a disk are reassuringly insensitive to the choice of prescription.  However, disk
  outflows can be sensitive to the choice of method, even for the same effective viscous
  strength.  We also consider the effects of eddy diffusion in the evolution of an
  accretion disk and show that it can interestingly affect the composition of outflows.
\end{abstract}

\maketitle

\section{Introduction}

It is progress, of a sort, that the realism of numerical relativity simulations is now
limited primarily by the same physical and computational challenges as is that of their
Newtonian counterparts.  Surely among the greatest of these is the multiscale nature of
turbulent fluid flow; kinetic energy at the system size or unstable mode wavelength
cascades through an inertial range of smaller scales until finally dissipated into
internal energy at scales far below what can be captured numerically.  High Reynolds
numbers, and hence this turbulent cascade, are expected in all of the main
problems in nonvacuum numerical relativity:  binary neutron star mergers, black hole-neutron
star mergers, core-collapse supernovae, and collapsars.

One solution would be to pursue much higher resolution simulations.  The highest resolution
binary neutron star mergers have grid spacings of order twenty meters~\cite{Kiuchi2015}.  Meanwhile,
several groups are designing computational infrastructure that will allow scaling up to
hundreds of thousands of threads (e.g.~\cite{kidder:16}).  Another, complementary strategy,
is to model subgrid-scale
transport effects by adding effective stress, heat and particle conduction, and dynamo terms
to the large-scale evolution equations.  Several such attempts have been made for
relativistic hydrodynamics in the context of binary neutron star post-merger
remnants~\cite{duez:04,Giacomazzo:2014qba,Shibata:2017xht,Fujibayashi:2017puw,Radice:2017}
and black hole accretion (also often with post-merger applications)~\cite{Sadowski:2014awa,
  Fujibayashi:2020qda}
and of course in Newtonian hydrodynamics
turbulence modeling is a vast endeavor.  (For book-length treatments, see~\cite{wilcox2006turbulence,
  pope2000turbulence}.  Of particular
interest to relativistic astrophysics is the incorporation of turbulence effects on core collapse
supernovae that might not be directly captured due to resolution limits or dimensional
reduction~\cite{murphy:11,couch:15a,radice:15a,Mabanta18,Couch:2019mrd}.)
They have the advantage that grids can remain small and simulations cheap, so that parameter
explorations can readily be carried out.  On the other hand, to be believable, the added
terms must be calibrated to and validated by expensive high-resolution simulations.  Probably,
both strategies will play a role in the successful exploration of turbulent-fluid dynamical-spacetime
systems.

An important distinction should be made among subgrid models.  (On this distinction, see
e.g.~\cite{Wyngaard:2004}).
In what we will call ``large-eddy simulations'', it is assumed that a significant portion of the
inertial range is resolved, and subgrid stress terms are computed as an extrapolation of the
character of resolved turbulence to subgrid scales (e.g.~\cite{smagorinsky1963general,
  bardina1980improved,grete2015nonlinear,Grete:2017fyi}).  An example of such methods
is the gradient model which has recently been adapted to relativistic magnetohydrodynamics
by Carrasco, Vigan{\`o}, and Palenzuela~\cite{Carrasco:2019uzl} (see also~\cite{Vigano:2019}).
The subgrid dynamo term of Giacomazzo~{\it et al}~\cite{Giacomazzo:2014qba} might also fit
into this category, because the field growth is stopped when the magnetic energy density
approaches an estimate of the subgrid turbulent kinetic energy density.   Alternatively, one
may not resolve the turbulence at all (or fail to model the physics that inputs energy into
the turbulent cascade).  In this case, subgrid stresses must be assigned as functions of the
resolved laminar flow, and one has a mean-field model.  (It is common also to introduce new
evolution variables in the large-scale evolution representing, for example, the turbulent
kinetic energy.)  In this paper, we shall mostly be
concerned with mean-field models.  The most famous is the alpha-viscosity prescription of
Shakura and Sunyaev~\cite{shakura:1973}, and a number of the above-mentioned numerical
relativity
studies~\cite{duez:04,Shibata:2017xht,Fujibayashi:2017puw,Fujibayashi:2020qda}
follow the alpha-viscosity
path of modeling unresolved turbulence as a viscosity via the Navier-Stokes equations.  It
should be remembered that momentum transport is only one of the large-scale effects of
turbulence.  By analogy with the kinetic theory of gases, one also expects turbulent
heat conduction, turbulent eddy diffusion of scalar quantities like composition variables,
and a turbulent effective pressure.

Subgrid transport has recently been introduced into binary neutron star merger simulations
by Radice~\cite{Radice:2017}.  As described in more detail below, Radice considers his added
stress
terms to be the results of an averaging procedure with an imposed closure which is similar
to but not the same (nor intended to be the same) as the relativistic Navier-Stokes equations. 
Because his specification of the mixing length does not rely on locally-measured turbulence (most recently, it is calibrated to high-resolution MHD merger simulations~\cite{Kiuchi_2018}), the resulting
model is a mean-field model (a calibrated one) by our definition (although it could be extended into a large eddy
model by our definition by using local velocity gradients to estimate the effective viscosity,
as done by Smagorinsky and subsequent
large-eddy models~\cite{smagorinsky1963general}), and so it can be compared to
the Navier-Stokes simulations of the Illinois and SACRA groups~\cite{duez:04,Shibata:2017xht}.

In this paper, we investigate both Navier-Stokes and Radice-style momentum transport models.
Our formulations differ from some others in the literature mentioned above in that we retain the
same evolution variables as in ideal hydrodynamics, so that the recovery of primitive variables
is independent of our various non-ideal transport modifications.  
We study the proper formulation of both models and analyze their stability.  We perform
numerical experiments on both of the main configuration types that appear in numerical
relativity:  a differentially rotating compact star and a neutrino-cooled black hole accretion
disk.  Finally, we look at the effect of other types of turbulent mixing on a representative
accretion disk system.  In particular, we consider the effect of turbulent effective heat
flows on the mass of the outflow and the effect of eddy diffusion on its composition. 
The importance of the outflow mass is obvious, but the composition distribution of disk ejecta is
also an output of post-merger simulations of great importance for kilonova
predictions~\cite{Metzger2017}, where the relevant composition variable in this
case is the electron fraction $Y_e$.  The lack of turbulent composition mixing in prior studies that model transport by an
effective viscosity is potentially one of the major differences between these studies and
proper (but expensive) magnetohydrodynamic simulations which incorporate all effects of
turbulence.  (There are, of course, other major differences, including the effects of
a large-scale B field, which no local transport model will be able to capture.)

We find that, in mean-field momentum transport models, it is crucial to properly
include turbulent heating to the energy equation.  Failure to do so results in unphysical
behavior, most notably a nonaxisymmetric instability in rotating stars which often appears
only after the star has come close to uniform rotation (so that the effect of the transport
is presumed to be nearly done).  While simple mean field closure relations are not
four-dimensionally covariant,
for the types of problems (and coordinate systems) common to numerical relativity, the
difference from a Navier-Stokes evolution is quite modest.  The only exception, although
a very important one, is in the outflow mass, for which the differences can be quite
significant.  In our test case, heat diffusion significantly increases the mass of
ejected matter, while eddy diffusion can affect the peak and width of the $Y_e$ distribution.

The paper is organized as follows.  In section~\ref{sec:NS}, we derive the relativistic
Navier-Stokes equations, put them in a convenient form for numerical implementation, and
analyze their stability.  In section~\ref{sec:TMS}, we work from a framework of averaging
the effects of subgrid stresses, looking particularly at the treatment of the energy
equation.  In section~\ref{sec:startest}, we test our transport methods on a differentially
rotating star problem, first looking at the early-time evolution of the rotation profile
and entropy, then at the long-term stability issues.  In section~\ref{sec:disktest}, we present
numerical experiments on a black hole accretion system.  We summarize our findings in section~\ref{sec:conclusion}.

\section{The Relativistic Navier-Stokes Equations}
\label{sec:NS}

\subsection{Metric and Fluid Variables}

In the 3+1 formalism, spacetime is foliated into spacelike hypersurfaces $\Sigma(t)$ parametrized by the timelike
coordinate $t$. The spacetime metric is $g_{\mu \nu}$, which we decompose as
\beq
ds^2 = -\alpha^2 + \gamma_{ij} (dx^i + \beta^i) (dx^j + \beta^j)
\eeq
where $\alpha$ is the lapse, $\beta^i$ the shift and $\gamma_{ij}$ the 3-metric.
The unit normal $n^{\mu}$ to a slice $\Sigma$ is then
\beq
n^\mu = \frac{1}{\alpha} (t^\mu -\beta^\mu) = (1/\alpha,-\beta^i/\alpha)\ .
\eeq
The extrinsic curvature of a slice $\Sigma$ is defined as
\beq
K_{\mu\nu} = -\nabla_\nu n_\mu -n_\nu \gamma_\mu^\lambda \nabla_\lambda (\ln \alpha) = -\frac{1}{2}\mathcal{L}_n \gamma_{\mu\nu}
\eeq
where $\mathcal{L}_n$ is the Lie derivative along the normal $n^\mu$.

For a perfect fluid, we define the stress-energy tensor of matter as
\beq
\label{eq:Tideal}
T^{\mu \nu} = \rho_0 h u^\mu u^\nu + P g^{\mu \nu}
\eeq
where $\rho_0$ is the baryon density, $h=1+P/\rho_0 + \epsilon$ is the specific enthalpy,
$P$ is the pressure and $\epsilon$ the specific internal energy.
The 4-velocity $u^\mu$ can be decomposed in 3+1 form
\beq
u^\mu = W (n^\mu + v^\mu)
\eeq
where $W$ is the Lorentz factor, and $v^\mu$ the 3-velocity. Note the we require $v^\mu n_\mu=0$ (i.e. $v^t=0$).
In components, we have
\beqn
u^\mu &=& (W/\alpha, W[v^i - \beta^i/\alpha])\\
u_\mu &=& (W[-\alpha+\beta^iv_i], Wv_i).
\eeqn
The conservative variables used for numerical evolutions are
\beqn
\rho_* &=& \rho_0 W \sqrt{\gamma}\\
\label{eq:tau_general}
\tau &=& \sqrt{\gamma}n_\mu n_\nu T^{\mu \nu} - \rho_*\\
\label{eq:s_general}
S_i &=& -\sqrt{\gamma} n_\mu \gamma_{i \nu} T^{\mu \nu}.
\eeqn
Although not an evolution variable, the purely spatial projection of
the stress tensor appears in source terms.
\begin{equation}
  \label{eq:sij_general}
  S_{ij} \equiv \alpha\sqrt{\gamma}\gamma^{\alpha}_i\gamma^{\beta}_jT_{\alpha\beta}
\end{equation}
For a perfect fluid, Eq~\eqref{eq:Tideal}, the stress tensor projections are
\beqn
\label{eq:tau_ideal}
\tau &=& \rho_* (hW -1) - P \sqrt{\gamma} \\
\label{eq:s_ideal}
S_i &=& \rho_* h u_i .
\eeqn

The evolution equations are the conservation of baryon number and the projections of the Bianchi identity $\nabla_\mu T^{\mu \nu}=0$:
\beqn
\nabla_\mu (\rho_0 u^\mu) &=& 0\\
\nabla_\mu (T^{\mu \nu} n_\nu) &=& T^{\mu \nu} \nabla_\mu n_\nu\\
\nabla_\mu (T^{\mu \nu} g_{\nu i}) &=& 0\ ,
\eeqn
which can be expanded as
\beqn
\partial_t \rho_* &+& \partial_i (\rho_* v_T^i) = 0\\
\partial_t \tau &+& \partial_i (\tau v_T^i+P\sqrt{\gamma} \alpha v^i) = \nonumber\\
\label{eq:tausource-ideal}
&&\alpha P K \sqrt{\gamma} - S^i \partial_i \alpha + S^iS^jK_{ij} \frac{\alpha}{\rho_* Wh}\\
\label{eq:ssource-ideal}
\partial_t S_j &+& \partial_i (S_j v_T^i + \alpha P\sqrt{\gamma} \delta^i_j) = S_i \partial_j \beta^i\\
&& +\sqrt{\gamma}P (\partial_j \alpha + \frac{\alpha \partial_j \gamma}{2\gamma}) - \rho_* Wh \partial_j \alpha + \frac{\alpha S^i S^k \partial_j \gamma_{ik}}{2\rho_* Wh} \nonumber
\eeqn
where we defined the transport velocity
\beq
v_T^i = \frac{u^i}{u^t} = \alpha v^i - \beta^i.
\eeq
For a general stress tensor, the source term for $\tau$ is $-\alpha \sqrt{\gamma} T^{\mu \nu} \nabla_\nu n_\mu = -S^i\partial_i\alpha + S_{ij}K^{ij}$, and the source term for $S_i$ is $\alpha \sqrt{\gamma} T^{\mu \nu} \partial_i g_{\mu \nu}/2$.

\subsection{The Shear Tensor}

A simple prescription for a viscous fluid is to include a shear viscosity but no bulk viscosity. Then, the stress-energy tensor becomes
\beq
T^{\mu \nu} = T^{\mu \nu}_{\rm ideal} + \tau^{\mu\nu}
= T^{\mu \nu}_{\rm ideal} - 2\eta \sigma^{\mu \nu}
\eeq
where
\beq
\sigma_{\mu \nu} = \nabla_{(\mu}u_{\nu)} + u^\alpha \left(\nabla_\alpha u_{(\mu}\right) u_{\nu)}-\frac{1}{3} \nabla_\alpha u^\alpha h_{\mu \nu}
\eeq
is the shear tensor and
\beq
h_{\mu \nu} = g_{\mu \nu} + u_\mu u_\nu.
\eeq
The coefficient $\eta$ sets the strength of the viscosity.  For a physical
viscosity, kinetic theory would lead one to expect $\eta \approx \rho_0c_s\ell$,
where $c_s$ is the sound speed and $\ell$ is the mean free path of the
constituent particle.

We can take advantage of the identity $\sigma^{\mu \nu} u_\mu = 0$ to write
\beqn
\sigma_{jt} &=& -\sigma_{ij} v_T^i\\
\sigma_{tt} &=& \sigma_{ij} v_T^i v_T^j\\
\sigma^t_i &=& \sigma_{ij}\frac{v^j}{\alpha}\\
\label{eq:sigij}
\sigma^i_j &=& \sigma_{jk} \left(\gamma^{ik} - \frac{\beta^i v^k}{\alpha}\right)\\
\sigma^t_t &=& -\sigma_{ij} \frac{v^i_T v^j}{\alpha}\\
\sigma^{tt} &=& \frac{\sigma_{ij}}{\alpha^2}v^iv^j\\
\label{eq:sigti}
\sigma^{ti} &=& \left(\frac{\gamma^{ik}}{\alpha}-\frac{\beta^iv^k}{\alpha^2}\right)\sigma_{kj} v^j\\
\sigma^{ij} &=& \sigma_{lm} \left(\gamma^{il}\gamma^{jm} + \frac{\beta^i\beta^j}{\alpha^2}v^lv^m\right)\nonumber\\
&&-  \sigma_{lm}\left(\frac{\beta^i}{\alpha}\gamma^{jl}v^m +  \frac{\beta^j}{\alpha}\gamma^{il}v^m\right),
\eeqn
and we only need to compute $\sigma_{ij}$ to recover the full 4-dimensional tensor.  This 3D tensor can be simplified as
\beqn
\sigma_{ij} &=&\frac{W}{2\alpha} \left( u_j \partial_t u_i + u_i \partial_t u_j\right) -  \frac{h_{ij}}{3 \alpha \sqrt{\gamma}} \partial_t (\sqrt{\gamma} W)  \nonumber\\
&+& \frac{1}{2}\left( \partial_i u_j +  \partial_j u_i \right) - \frac{h_{ij}}{3 \alpha \sqrt{\gamma}} \partial_k  (W \sqrt{\gamma}v^k_T)  \nonumber\\
&-& W (K_{ij} + v_k \Gamma^{(3)k}_{ij}) - \frac{W}{2\alpha} (u_i \partial_j u_t + u_j \partial_i u_t) \label{eq:sigma} \\
&+& \frac{W v^k_T u_i}{2\alpha}(\partial_k u_j - \partial_j u_k)  + \frac{W v^k_T u_j}{2\alpha} (\partial_k u_i - \partial_i u_k) \nonumber
\eeqn
where $\Gamma^{(3)k}_{ij}$ are the 3-dimension Christoffel symbols associated with $\gamma_{ij}$.
The above expression only requires time derivatives of $u_i$ and $(\sqrt{\gamma}W)$.

\subsection{Evolution Equations}

The inclusion of the viscous shear tensor induces modification
to the evolution equation for $\tau$ and $S_i$. The fluxes are now
\beqn
\label{eq:fluxtau}
F^i_\tau &=& \tau v^i_T + P \alpha \sqrt{\gamma} v^i - 2\eta \alpha^2 \sqrt{\gamma} \sigma^{ti}\\
\label{eq:fluxS}
F^i_{S_j} &=& S_j v_T^i + \alpha P\sqrt{\gamma} \delta^i_j - 2 \eta \alpha \sqrt{\gamma} \sigma_j^i
\eeqn
and the source terms
\beqn
S_\tau &=&  S_\tau^{\rm ideal} - 2\eta \alpha \sqrt{\gamma} (\sigma_{ij}K^{ij} - \sigma^{ti} \partial_i \alpha) \nonumber\\
&+& \partial_t (2\eta \alpha^2 \sqrt{\gamma} \sigma^{tt}) \\
S_{S_j} &=& S_{S_j}^{\rm ideal} - \eta \alpha \sqrt{\gamma} \sigma^{\mu \nu} \partial_j g_{\mu \nu} \nonumber\\
&+& \partial_t (2\eta \alpha \sqrt{\gamma} \sigma_j^t)\ ,
\eeqn
where $S_\tau^{\rm ideal}$ and $S_{S_j}^{\rm ideal}$ are the source terms for ideal hydrodynamcs, given as the right-hand sides of equations~\eqref{eq:tausource-ideal} and~\eqref{eq:ssource-ideal}, respectively.

\subsection{Stability}
\label{sec:stability}

We will consider stability on a flat background, for perturbation around a homogeneous fluid configuration.
The perturbations will be planar waves proportional to $\exp{(\Gamma t + i k x)}$.  
First order theories of the Navier-Stokes equation, including the equations derived in the previous section, are acausal and have 
unstable modes with extremely rapid growth rate~\cite{Hiscock:1983zz}. One can however recover a stable and covariant
viscous formalism by going to second-order methods (Israel-Stewart viscosity~\cite{Israel:1976tn,ISRAEL1979341,Hiscock:1983zz}), which treat the viscous stress-tensor as an evolved
variable.

Suppose that at a particular instant in time, the viscous coefficient $\eta$ is
a function of the local density $\rho_0$ and temperature $T$, while the shear tensor is
a function of $u^{\mu}$ and its derivatives.  Then the instantaneous values of
$-2\eta\sigma_{ij}$ at event $x^{\mu}$ are
\begin{equation}
\tau_{ij}^{\rm inst}(x^{\mu}) = -2\eta(\rho_0(x^{\mu}),T(x^{\mu}))\sigma_{ij}(u^{\alpha}(x^{\mu}),\partial_{\beta}u^{\alpha}(x^{\mu}))\nonumber\ ,
\end{equation}
Instead of setting the viscous stress tensor
$\tau_{ij}$ at each event $x^{\mu}$ equal to $\tau_{ij}^{\rm inst}(x^{\mu})$,
we evolve $\tau_{ij}$ according to
\begin{equation}
  \label{eq:tauvisc_ev}
\partial_t\tau_{ij} = -\frac{1}{t_d}\left(\tau_{ij}-\tau_{ij}^{\rm inst}\right)
\end{equation}
or
\begin{equation}
  \label{eq:tauvisc_ev2}
\mathcal{L}_{\vec{t}+\vec{v_T}}\tau_{ij} = -\frac{1}{t_d}\left(\tau_{ij}-\tau_{ij}^{\rm inst}\right)
\end{equation}
The advection term in the second version is probably preferable for systems with
high velocities (at least, high velocities not along a symmetry of the fluid configuration). 
Alternatively, if $t_d$ is small compared to physical timescales, the driver can handle the
advection itself.  In the applications in this paper, systems are mostly axisymmetric and
velocities mostly azimuthal, so we find better performance with Eq.~\eqref{eq:tauvisc_ev}.

Consider the case of a perturbation propagating along the direction of the
fluid motion ($u_y=u_z=0$, $u_x\neq0$).  We perturb the energy and momentum
equations and close the system by specifying an equation of state
$\epsilon(\rho_0,T)$, $P(\rho_0,T)$.  Consider
transverse modes (involving $\delta u_A$, $\delta \tau_{xA}$), where
capital roman letters will stand for indices $y,z$.  We have the constraint
\beq
\delta \tau^{t\mu} = \frac{u_x}{W} \delta \tau^{x\mu}.
\eeq

The perturbed momentum equation along $y,z$ becomes 
\beq
\rho_0 h (\Gamma W + ik u_x)\delta u_A  + (\Gamma \frac{u_x}{W} +ik ) \delta \tau_{xA} = 0 \nonumber
\eeq
and the evolution equations for $\delta \tau_{xA}$ without advection term is
\beq
\label{eq:tau_ev_without_advection}
(\Gamma t_d + 1) \delta \tau_{xA} = -\eta W (ik W + u_x \Gamma ) \delta u_A.
\eeq
With the advection term, it would be
\beq
\label{eq:tau_ev_with_advection}
(\Gamma t_d + 1) \delta \tau_{xA} = -\eta W (ik W + u_x \Gamma ) \delta u_A
- \frac{u_x}{W}t_dik\delta\tau_{xA}.
\eeq

Concentrating for the moment on the system without advection of $\tau_{xA}$
(Eq.~\eqref{eq:tau_ev_without_advection}), we thus have the system
\beq
\label{eq:matrix}
\begin{pmatrix} 
\rho_0 h (\Gamma W + ik u_x) & \frac{1}{W} (\Gamma u_x +ikW) \\ 
\eta W (ik W + u_x \Gamma ) & 1 + \Gamma t_d
\end{pmatrix} 
\begin{pmatrix}
\delta u_A \\  \delta \tau_{xA} \end{pmatrix} 
= \begin{pmatrix} 0 \\ 0 \end{pmatrix}
\eeq

Taking the determinant of the matrix and setting it equal to zero, we solve
for $\Gamma$ and find
\beq
\Gamma = \frac{-\rho_0 h W + i ( \cdots ) \pm \Delta^{1/2}}{\cdots}\ ,
\eeq
where we have neglected to expand factors not relevant to the question of
stability, which requires only $\Re{(\Gamma)}<0$.  In the above,
\beqn
\label{eq:delta}
\Delta &=& (\rho_0 h W)^2 \nonumber \\
&-&2ik u_x \rho_0 h \left( \rho_0 h Wt_d + 2 \eta \right) \nonumber \\
&-& k^2 \left( (\rho_0 h u_x t_d)^2 + 4\eta W \rho_0 h t_d\right).
\eeqn
For stability, we need $\Re{\left(\Delta^{1/2}\right)}\leq \rho_0 h W$. 
Note that the marginal stability case $\Re{\Delta^{1/2}} = \Re{\Delta_{\rm ms}^{1/2}}=\rho_0 h W$
is of the form $\Delta_{\rm ms}=(\rho_0 h W \pm ikB)^2 = (\rho_0 h W)^2 \pm 2ik \rho_0 h W B - k^2 B^2$
for some $B\in\mathbb{R}$.  On the other hand, Eq.~\eqref{eq:delta} has the form
$\Delta = (\rho_0 h W)^2 - 2ik\rho_0 h W C - k^2 D$ for $C$,$D\in\mathbb{R}$.  Thus, $D=C^2$ is the condition
for marginal stability, while $D>C^2$ is the condition for stability, which can be written
\beq
(\rho_0 h u_x t_d)^2 + 4\eta W \rho_0 h t_d > \frac{u_x^2}{W^2} ( \rho_0 h Wt_d + 2 \eta)^2
\eeq
or
\beq
\label{eq:stability_condition}
 t_d > \frac{W}{\rho_0 h} \left( \eta \frac{W^2-1}{W^2}\right)\ ,
\eeq
which shows that $t_d>0$ is required whenever $\eta\ne 0$.

One can repeat the above analysis with the advection term and arrive at exactly
the same condition.

\subsection{Implementation}

The Israel-Stewart formulation of the Navier-Stokes equations 
can be implemented in a numerical relativity code in
either of two ways.  First, one can retain the general definition of $\tau$
and $S_i$ in terms of the total stress tensor, Eq.~\eqref{eq:tau_general}
and~\eqref{eq:s_general}, which would then include viscous terms.  In this
case, the algorithm for recovering primitive variables must be altered. 
This is the method chosen by Fujibayashi {\it et al}~\cite{Fujibayashi:2017puw}.
Second, one could retain the definition of $\tau$ and $S_i$ in terms of
fluid variables, Eq.~\eqref{eq:tau_ideal} and~\eqref{eq:s_ideal} in which case
all terms from the divergence of the viscous stress tensor are regarded as
source or flux terms.  In this case, primitive variable recovery is not affected
by viscosity, but the viscous source terms require knowing the time
derivatives $\partial_t(\sqrt{\gamma}W)$ and $\partial_tu_i$.  We choose
this second method.  The needed time derivatives are estimated by storing
$\sqrt{\gamma}W$ and $u_i$ at the previous timestep and then computing at
each step $k$ the backward-centered time derivative
$\partial_tX(t_k) = [X(t_k) - X(t_{k-1})]/(t_k-t_{k-1})$.

Second-order theories of viscosity also require the evolution of the stress
tensor.  We promote $A_{ij}\equiv -2\eta\sigma_{ij}W\sqrt{\gamma}$ to
be a new evolved variable with evolution equation
\begin{equation}
  \label{eq:driverA}
  \partial_tA_{ij} + \mathcal{A}_{ij} = -\frac{A_{ij}-A_{ij}^{\rm inst}}{t_du^0}
\end{equation}
where $\mathcal{A}_{ij}$ is an optional advective term $\mathcal{L}_{v_T}A_{ij}$.

We evolve Eq.~\eqref{eq:tauvisc_ev} using
implicit time steps.  This can be written clearly if we momentarily
suppress indices and use subscripts for timesteps, so that $A_k$
is a component of $A_{ij}$ at step $k$.  Given $A_{k-1}$ and
$A^{\rm inst}_k$, we take a step of size $\Delta t=t_k-t_{k-1}$ by
\begin{equation}
  A_k = \frac{A_{k-1} + A^{\rm inst}_k\Delta t /(u^0t_d) - \mathcal{A}(A_{k-1})\Delta t}{1 + \Delta t/(u^0t_d) }
\end{equation}
We have experimented with the choice of $t_d$, setting it to some fraction
of the dynamical timescale--a constant for star runs, a fraction of the
local Keplerian period for disk runs.  We find our results to be insensitive
to its value so long as $t_d$ is small compared to the dynamical timescale
and the instability is not triggered (cf. Eq.~\eqref{eq:stability_condition}).

We add an option to suppress viscosity at low densities or very close to
the black hole, anticipating the possibility of excess artificial viscous
heating in these regions.  From Eq.~\eqref{eq:stability_condition}, the
minimum $t_d$ scales with $\eta$, so if we suppress $\eta$ by some
function of density or distance from the black hole, we reduce $t_d$ by
the same factor.

\subsection{Other Covariant Implementations}

The procedure of promoting viscous stress to an independent evolved
variable can be extended to other dissipative fluxes (heat flux, etc),
a process done systematically in the field of extended irreversible
thermodynamics~\cite{Jou:1988}.  However, even the Israel-Stewart
theory is known to incorrectly handle strong
shocks~\cite{Olson:1991pf,Geroch:1991},
which could be a serious problem in some numerical relativity
applications.  A different way of improving the behavior of the
relativistic Navier-Stokes equations, explored by
Lichnerowicz~\cite{Lichnerowicz:1955}
and Disconzi~\cite{disconzi:14}, is to replace $u_i$ by $h u_i$ in the formula
for the shear tensor.

Recently, a new formulation of relativistic viscous hydrodynamics
has been introduced by Bemfica, Disconzi, Noronha, and Kovtun
(BDNK)~\cite{Bemfica:2017wps,Bemfica:2019knx,Bemfica:2020zjp}.
They begin from a general expansion of the stress tensor
in covariant terms with first derivatives of fluid variables.  Each
term will be multiplied by a coefficient, and BDNK find conditions on
the coefficients that guarantee stability and causality.  Kinetic
theory determines some coefficients, but freedom remains in the choice
of others, different values representing different choices
in the definition of fluid variables out of equilibrium.  Presumably,
a similar procedure could be followed to model turbulent effective
stresses, with similar freedoms reflecting different ways to
average over small-scale eddies.  However, the BDNK formulation has
not yet been used in numerical relativity.

In this paper, we restrict attention to momentum transport methods
already in use in numerical relativity, which is already enough to
get some indication of how sensitive simulations are likely to be
to the choice of method.

\section{Effective Reynolds Stress from Subgrid-Scale Turbulence}
\label{sec:TMS}

\subsection{Filtered Variables and Evolution Equations}

As was pointed out by Boussinesq and Prandtl over a century ago,
the mixing of momentum by turbulent eddies is analogous to molecular
transport in a gas, suggesting that on scales much larger than the
eddies, the mean stress from turbulence might be like a viscosity. Once
again, we would expect $\eta_T \approx \rho_0c_s\ell$, but now $\ell$ is
the {\it mixing length} associated with the turbulence.

To pursue the kinetic theory analogy, divide the velocity flow into
large and small scales: $v^i = \overline{v}^i + \delta v^i$.  The
two components are defined by an averaging / low-pass filtering
operator $\langle\cdots\rangle$, such that
$\langle v^i \rangle = \overline{v}^i$, $\langle\delta v^i\rangle=0$.
Then one applies the filter to the ideal energy and momentum equations
to obtain evolution equations for $\overline{\tau}$ and $\overline{S}_i$.

The subleties that arise can be illustrated in the case of Minkowski
spacetime and incompressible small-scale turbulence.  Then the filtered
equations can be written
\begin{eqnarray}
  \nonumber
  \partial_t\overline{S}_i + \partial_j(\overline{S_iv^j}
  +\delta_i^jP) &=& 0 \\
  \nonumber
  \partial_t\overline{\tau}
  + \partial_j[\rho_0h\overline{Wv^j}-\rho\overline{v}^j]
  &=& 0
\end{eqnarray}
where $\overline{S}_i=\rho h\overline{Wv_i}$.  Note that
$\overline{S_iv^j}\ne\overline{S}_i\overline{v}^j$ and
$\overline{Wv_i}\ne\overline{W}\overline{v}_i$.  Assuming
$\delta v^i$ is not highly relativistic, we can Taylor expand
the Lorentz factor in $\delta v^i$
and get $\overline{W}=W(\overline{v}^i)$,
$\delta W=\overline{W}^3\overline{v}^j\delta v_j$.  Then
\begin{equation}
  \langle v^iW\rangle = \overline{W}\overline{v}^i
  + \overline{W}^3\overline{v}^j\left\langle\delta v_j\delta v^i\right\rangle\ ,
  \nonumber
\end{equation}
Similarly, we define
\begin{equation}
  \langle S_i v^j\rangle \equiv \overline{S}_i\overline{v}^j + \tau_i{}^j\ ,
\end{equation}
where we now {\it redefine} $\overline{S}_i$ to be
$S_i(\overline{v}_i)\ne \langle S_i\rangle$.  This redefinition is desirable
because it preserves the relationship between primitive
and conservative variables at the filtered level.
One finds $\tau_i{}^j$ to be
\begin{equation}
  \tau_i{}^j = \rho_0h\overline{W}\langle\delta v_i\delta v^j\rangle
  + O(|\overline{v}|^2\times\langle\delta v\delta v\rangle)
  \nonumber
\end{equation}
with the omitted terms coming from correlation between $\delta W$ and
$\delta v$.  To second order in $\overline{v}^i$, $\tau_{ij}$, we have
\begin{eqnarray}
  \partial_t \overline{S}_i + \partial_j(\overline{S}_i\overline{v}^j+\tau_i{}^j)
  &=& 0 \\
  \partial_t \overline{\tau}
  + \partial_j(\overline{S}^j + \overline{v}^k\tau_k{}^j) &=& 0
\end{eqnarray}
Note that the extra term in the energy equation, from the difference between
the redefined $\overline{S}_i$ and $\langle S_i\rangle$, is necessary to
correctly recover the Newtonian limit, and in particular to secure energy
conservation in this limit.  One could handle this instead by adding the
$\tau\cdot v$ term to the definition of the conservative variable $\overline{S}_i$,
i.e. by keeping the definition $\overline{S}_i = \langle S_i\rangle$, but
we find it more straightforward to retain the standard relations between
primitive and conservative variables.

Returning to the case of general metric and relativistic mean velocities,
the cleanest way to obtain the transport terms proportional to the mean
velocity is to impose that $\tau_{ij}$ be the spatial components of a
4D tensor that obey the usual orthogonality conditions for a shear tensor:
$\tau_{\mu\nu}u^\nu=0$.  Then the added terms to the energy and momentum
flux are the same as in equations~\eqref{eq:fluxS} and~\eqref{eq:fluxtau}, with
$\tau^{\alpha\beta}=-2\eta\sigma^{\alpha\beta}$ and with $\sigma^i_j$ and
$\sigma^{ti}$ related to $\sigma_{ij}$ as in Eq.~\eqref{eq:sigij}
and~\eqref{eq:sigti}.

To test the effect of the velocity-dependent terms, we write these terms as
\begin{eqnarray}
  \label{eq:turb-visc-dtau}
  \partial_t\tau &=& \cdots
  - \partial_j[(k_1\alpha\gamma^{jk}-k_2\beta^jv^k)v^l\tau_{kl}] \\
  \nonumber
  & & + \alpha\sqrt{\gamma}\gamma^{ki}\gamma^{lj}K_{kl}\tau_{ij}  \\ 
  \label{eq:turb-visc-dS}
  \partial_tS_i &=& \cdots
  - \partial_j[\sqrt{\gamma}\tau_{im}(\alpha\gamma^{mj}-k_2\beta^jv^m)] \\
  & &
\nonumber
  + \frac{1}{2}\sqrt{\gamma}\tau_{jk}(-2k_2v^jv^k\partial_i\alpha
  + 2k_2v^k\partial_i\beta^j-\alpha\partial_i\gamma^{jk})\ ,
\end{eqnarray}
where ``$\cdots$'' indicates the perfect fluid terms, and $k_1$, $k_2$ are
constants.  To enforce $\tau_{\mu\nu}u^\nu=0$, these constants should be
$k_1=k_2=1$.  $k_1=1$, $k_2=0$
has only the term needed to recover the Newtonian limit.  $k_1=k_2=0$ would
be to take equations (8) and (9) from~\cite{Radice:2017} while not accounting
for the difference between $S_i(\langle v_i\rangle)$ and $\langle S_i\rangle$
(which is the formalism used in~\cite{Radice:2017,radice2020binary}).

One can carry out a stability analysis as in Section~\ref{sec:stability}
for the above mean-field turbulence theory.  Because $\tau$ no longer appears
under the time derivative in the left-hand side of the momentum equation, the
$\Gamma u_x/W$ term in the upper right hand entry of the matrix in
Eq.~\eqref{eq:matrix} is no longer
present.  Anticipating that $t_d>0$ will no longer be required, we can set
$t_d=0$ and solve a linear equation for $\Gamma$, finding the real part to
be unconditionally negative.

\subsection{The Closure Condition}

Guided by the Smagorinsky closure~\cite{smagorinsky1963general} of
Newtonian turbulence modeling,
Radice~\cite{Radice:2017} proposes the following relativistic closure
\begin{equation}
  \tau_{ij} = -2\nu\rho_0hW^2\left[\frac{1}{2}(\nabla_i\overline{v}_j
    + \nabla_j\overline{v}_i)-\frac{1}{3}\nabla_k\overline{v}^k\gamma_{ij}\right]
  \ ,
  \label{eq:closure}
\end{equation}
where $\nabla_i$ is the 3D covariant derivative compatible with $\gamma_{ij}$. 

In a large-eddy simulation (as defined above), $\nu$ would be set by the
local state of the turbulence as determined by difference operators on the
smallest resolved scales.  This requires the simulation to resolve
some of the inertial range of the turbulence.  A mean field model is
needed if the turbulence is totally unresolved or if the physics driving
the turbulence is missing in the simulation.  For example, if turbulence
is driven by the magnetorotational instability and one's simulation does
not include magnetic fields, one would need a mean field closure condition
even if the hypothetical MRI wavelength is resolved.  The natural choice
is
\begin{equation}
  \nu = \ell c_s
\end{equation}
with $\ell$ the mixing length.

The evolution equations with the above closure are not--and are not meant
to be--exactly equivalent to the Navier-Stokes equations.  Nevertheless,
Eq.~\eqref{eq:closure} clearly does closely resemble a viscous stress
and behaves in a similar way.

The model is completed by choosing the mixing length $\ell$.  This will depend
on the system; the focus of Radice's work was binary neutron stars.  In his original
paper, he used constant values of $\ell$ set to be similar to the wavelength
of the fastest growing MRI mode for $B\sim 10^{14}$--$10^{15}$G.  More
recently~\cite{radice2020binary}, he has used a density-dependent $\ell(\rho_0)$, where the
function $\ell(\rho_0)$ was fit to the results of high-resolution MHD
simulations of binary neutron star mergers by Kiuchi~{\it et al}~\cite{Kiuchi_2018}.

\subsection{The Issue of Covariance}
\label{sec:covariance}

Equation~\eqref{eq:closure} is covariant with respect to spatial coordinate
transformations, but not with regard to general spacetime coordinate
transformations, a point made by Radice himself.  The filtering operator
is itself frame/slicing dependent, so we should not expect general covariance
in the final equations~\footnote{Unless, of course, one were to explicitly
add information about the filtering frame to the equations.  Any equations
can be put in generally covariant form given enough auxiliary variables.}. 
However, ``not covariant'' does not necessarily mean ``not valid''. 
Similar points are made by Carrasco~{\it et al}~\cite{Carrasco:2019uzl} with regard
to their relativistic subgrid code.  Since theirs is a large-eddy code,
they also point out that discretization for finite differencing itself violates
covariance in the same way and to a similar degree.  Also, covariance is
regained in their case, but not in the mean field case, in the limit of
infinite resolution, albeit trivially so because the subgrid terms then
disappear.

However, a non-covariant choice of closure may leave coordinate-independent
artifacts.  For example, one physically expects that, when the radius of
curvature is larger than the mixing length, momentum transport should
operate only when there is a nonzero shear as measured in a local Lorentz
frame, and heating should occur if and only if
$\sigma^{\alpha\beta}\sigma_{\alpha\beta}\ne 0$.  This is guaranteed for
the relativistic Navier-Stokes equations but not for Eq.~\eqref{eq:closure}.

Hereafter, we will refer to evolutions with the full Navier-Stokes equations,
stabilized by evolving the spatial stress tensor with the driver equation~\eqref{eq:driverA},
as ``Navier-Stokes'' or ``NS'' evolutions.  Mean-field turbulence evolutions using the
closure equation~\eqref{eq:closure} will be called ``turbulent mean stress'' or ``TMS'' evolutions.

\subsection{Diffusion of scalar quantities}

In addition to transporting momentum, turbulence leads to other effects that can be
understood qualitatively as transport by ``mixing''.  These include eddy diffusion of
particle species and turbulent heat transport.~\footnote{In some compressible turbulence models,
there is even a diffusion of density added to the continuity equation, although this
is sometimes avoided by using Favre rather than Reynolds average definition of the mean
velocity field~\cite{Favre_1992}.} 

We consider only the first of these effects.  We consider a scalar quantity $Y$,
say a species fraction that (up to reaction source terms) advects with the fluid,
so that $\rho_*Y$ obeys a continuity equation (possibly with reaction source terms).
Turbulent mixing will produce a flux of $\rho_*Y$ which we take to be
$F_{\rho Y}\approx \rho_0c_s\ell_D\nabla Y$, where we have given ourselves the freedom
of using a different mean free path for momentum transport and diffusion:
$\ell_D\equiv\lambda_D\ell$ for some constant $\lambda_D$.  Taking the divergence of this flux
(and ignoring the time derivative term), we get
\begin{equation}
  \partial_t(\rho_*Y) + \partial_i(\rho_*Yv_T^i) = \partial_j(\rho_*c_s\ell_D\partial^jY) + \cdots,
\end{equation}
where the ellipsis $\cdots$ indicates the other source terms.  

As with the TMS stress, this flux is not 4-dimensionally covariant.  A covariant treatment
would be, for example,
\begin{eqnarray}
  \nabla_{\mu}\mathcal{Y}^{\mu} &=& 0 \\
  \mathcal{Y}^{\mu} &=& \rho_0 Y u^{\mu} - \rho_0c_s\ell_D(g^{\mu\nu}+u^{\mu}u^{\nu})\nabla_{\nu}Y
\end{eqnarray}
as can easily be seen by going into a comoving local Lorentz frame.

Heat transport might be modeled in a similar way.  Since eddies (except near the dissipation
scale) evolve adiabatically, specific entropy $s$ rather than temperature would seem to be the
more appropriate scalar quantity to diffuse.  This would be in keeping with the normal practice
in mixing length theory treatments of convective stars (e.g.~\cite{Ruediger:1989}),
although here there is no presumption that eddies are buoyancy driven.  This could be captured
by a turbulent mean heat flux $q_i = \rho_0Tc_s\ell_S\nabla_is$ for some $\ell_S\equiv\lambda_S\ell$.
The corresponding covariant 4-vector obeying $q\cdot u = 0$ is
$q_{\mu}=\rho_0Tc_s\ell_S(\nabla_{\mu}s + u^{\nu}u_{\mu}\nabla_{\nu}s)$.  In the spirit of TMS, we
eliminate time derivatives by assuming entropy roughly advects $u^{\nu}\nabla_{\nu}s\approx 0$ (as
would be exactly true if it were a perfect fluid in the absence of shocks and radiation).  Then
\begin{eqnarray}
  q_{\alpha} &=& \rho_0Tc_s\ell_S (-v_T^i\nabla_is, \nabla_is) \\
  T^{\mu\nu}_{\rm heat} &=& q^{\mu}u^{\nu} + q^{\nu}u^{\mu} \\
  F_{\tau}^i &=& \cdots + \alpha W\sqrt{\gamma}(q^i + v_T^i q^t) \\
  F_{S_j}^i &=& \cdots + \alpha\sqrt{\gamma}(q^iu_j + q_ju^i) \\
  S_{\tau} &=& \cdots - S^i_{\rm heat}\partial_i\alpha + S^{ij}_{\rm heat}K_{ij} \\
  S_{S_j} &=& \cdots + \frac{1}{2}\alpha\sqrt{\gamma}T^{\mu\nu}_{\rm heat}\partial_ig_{\mu\nu}\ ,
\end{eqnarray}
where $S^i_{\rm heat}$ and $S^{ij}_{\rm heat}$ are projections of $T^{\mu\nu}_{\rm heat}$ as in
Eq.~\eqref{eq:s_general} and~\eqref{eq:sij_general}, and indices of the heat flux and 4-velocity are
raised and lowered using the 4-metric.

Simulations of magnetorotational turbulence find that the momentum transport is dominated by average Maxwell rather than average Reynolds stress, with the former around a few times larger than the latter (e.g.~\cite{Hawley:1995sy,Brandenburg:1995,Stone:1996}).  This suggests that $\lambda_D = \lambda_S = 1$ probaby overestimates mixing effects.  The choice of setting all mixing lengths equal, used at times below, is a useful way of checking what sort of influence turbulent particle diffusion and heat flux might have.
  
\section{Test on a differentially rotating star}
\label{sec:startest}

\subsection{Axisymmetric heating}

From an astrophysicists's point of view, turbulence is important primarily
for two reasons.  First, it transports angular momentum.  Second, it transfers
kinetic energy to small enough scales for it to be dissipated away as heat. 
Under the influence of a shear viscosity, a differentially rotating star will
approach uniform rotation on the viscous timescale $\sim R^2/\nu$, where $R$ is
the characteristic length of the shear flow, in this case the radius of the star.
The fluid will acquire entropy at a rate
\begin{equation}
  nT\frac{ds}{dt_{\rm prop}} = 2\eta \sigma_{\alpha\beta}\sigma^{\alpha\beta}\ ,
\end{equation}
where $n$, $T$, $s$, and $t_{\rm prop}$ are the number density, temperature,
specific entropy, and proper time along the fluid element, respectively.
For a Gamma-law equation of state $P=(\Gamma-1)\rho_0\epsilon$, $P(T=0)=\kappa\rho_0^{\Gamma}\equiv P_{\rm cold}$ this
can be written
\begin{eqnarray}
  nT\frac{ds}{dt_{\rm prop}} &=& \frac{P_{\rm cold}}{\Gamma-1}\frac{d}{d\tau}\left(\frac{P}{P_{\rm cold}}\right) \\ 
  \label{eq:heating}
  {\partial_{t}}({E}_{\ast}) &=&  -{\partial_{j}}({E}_{\ast}{v}_{T}^{j}) \\
  \nonumber
  &+& {\frac{\alpha\sqrt{\gamma}}{\Gamma}}
  (\frac{E_{\ast}}{W\sqrt{\gamma}})^{(1-\Gamma)} ({2\eta}{\sigma_{\alpha\beta}} {\sigma^{\alpha\beta}})
\end{eqnarray}
Where $E_{\ast} {\equiv} W\sqrt{\gamma}(\rho_{0}\epsilon)^{1/\Gamma} = W\sqrt{\gamma}(\frac{P}{\Gamma-1})^{1/\Gamma}$~\cite{baumgarteShapiroBook}.

\begin{figure}
\includegraphics[width=\columnwidth]{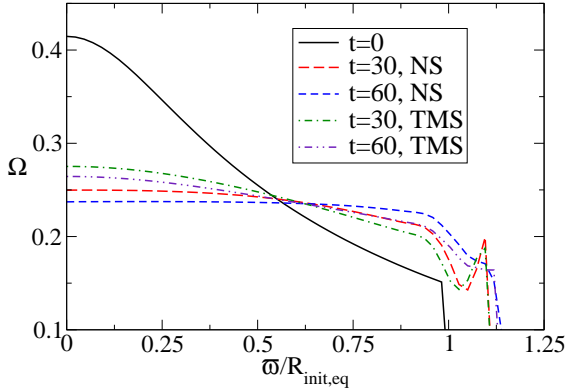}
  \caption{Equatorial angular velocity $\Omega$ at three times for NS and TMS evolution.  Angular velocity is shown as a function of coordinate cylindrical radius $\varpi$, normalized to the initial equatorial radius $R_{\rm init,eq}$.}
  \label{fig:Omega}
\end{figure}

\begin{figure}
\includegraphics[width=\columnwidth]{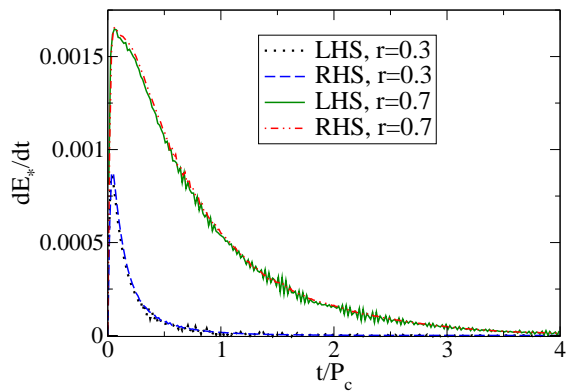}
  \caption{Time derivative of the specific entropy of two representative equatorial
    tracer particles [left and right hand sides of Eq.~\eqref{eq:heating}] for NS evolution.  Time $t$ is normalized to the initial central rotation period $P_c.$}
  \label{fig:NStracer}
\end{figure}

\begin{figure}
\includegraphics[width=\columnwidth]{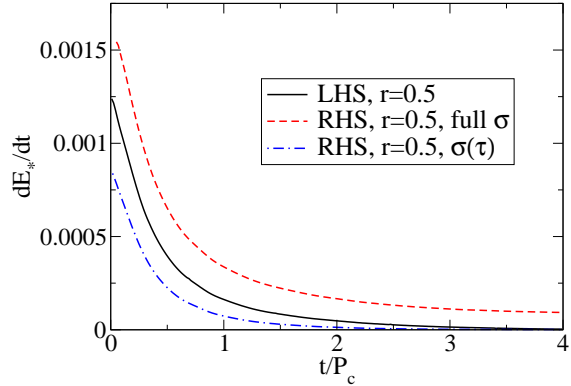}
  \caption{Numerical time derivative of the specific entropy of one representative equatorial
    tracer particle (left hand side of Eq.~\eqref{eq:heating}) for TMS evolution.  Also,
    estimates of the expected heating rate (right hand sides of Eq.~\eqref{eq:heating}) using
    the exact covariant shear tensor and using the TMS closure $\tau_{ij}$ (Eq~\eqref{eq:closure}).}
  \label{fig:TMStracer}
\end{figure}

As a first test of our momentum transport methods, we evolve a differentially
rotating relativistic star.  The initial equilibrium state is supplied by the
code of Cook, Shapiro, and Teukolsky~\cite{cook92}.  We use a polytropic equation of
state $P=\kappa\rho_0^{\Gamma}$ with $\kappa=1$, $\Gamma=2$.  The differential
rotation law is
\begin{equation}
  u^tu_\phi = R_{\rm eq}^2A^2(\Omega_c-\Omega)\, 
\end{equation}
where $R_{\rm eq}$ is the equatorial coordinate radius, $\Omega$ is the angular
velocity, $\Omega_c$ is the angular velocity on the axis, and the
differential rotation parameter $A$ is set to 1.  The star has baryonic mass
0.1756$\kappa^{1/2}c^2G^{-3/2}$, ADM mass 0.1627$\kappa^{1/2}c^2G^{-3/2}$,
and angular momentum 0.01402$c^3\kappa G^{-2}$.  The polar to equatorial radius ratio is
0.75.  The equatorial coordinate radius is 0.885$\kappa^{1/2}G^{-1/2}$.

We first evolve the star on a 2D grid assuming axisymmetry using the
techniques described in our recent paper~\cite{Jesse:2020}.  We use
a 2D Cartesian grid, with vertical and radial cylindrical polar coordinates
$z$ and $\varpi$.  We use $G=c=\kappa=1$ units.  The grid has 360 points covering
$-2\leq z\leq 2$ and 260 points covering $0 < \varpi\leq 2.6$.

We set the viscous
coefficient to $\eta=0.1P$, where $P$ is the pressure.  For this test, we are uninterested in
low density behaviors (e.g. winds), so we add an exponential suppression
factor when $\rho_0$ is below $\rho_{\rm cut}=0.1\rho_{0 \rm max}$, where
$\rho_{0 \rm max}$ is the initial maximum rest density:
$\eta \rightarrow \eta e^{-(\rho_{\rm cut}/\rho_0)^4}$.  The driving timescale
$t_d$ for $\sigma_{ij}$ is set to 0.12, much shorter than the initial central
rotation period of 15.

In Figure~\ref{fig:Omega}, we plot the angular velocity profiles.  For both
the NS evolution and the TMS evolution, the
rotation profile flattens inside the star, as expected.  We do see that
for the TMS evolution, the profile settles with a slight
nonzero level in the high-density region.  We pointed out in Section~\ref{sec:covariance} above that this would be a possibility, so its occurrence is not too surprising.  Because it is not derived from the covariant stress tensor $\sigma$,
the TMS stress $\tau$ can be zero when $\sigma$ is nonzero, and vice versa, for a general metric.  Because it is coordinate-independent, the scalar $\sigma_{\alpha\beta}\sigma^{\alpha\beta}$, which we measure in the viscous heating tests below, is the most reliable measure of whether local shear is really present.

In Figures~\ref{fig:NStracer} and~\ref{fig:TMStracer}, we plot the heating
rate of a representative tracer particle, plotting the left and right-hand
sides of Eq.~\eqref{eq:heating}.  One could also plot the integrals of
each side over the entire star, but then numerical error would be dominated
by the thin numerically heated layer at the surface of the star.  This heating
is present even in the absence of TMS or NS transport and is mainly
due to numerical viscosity, and thus is not expected to be directly related
to the chosen subgrid viscosity model.

For the TMS evolution, we calculate $\sigma_{\alpha\beta}$ appearing in the
heating rate (Eq.~\eqref{eq:heating}) in two ways.  We compute the full
covariant $\sigma_{\alpha\beta}$ that would appear in the Navier-Stokes
equations (Eq.~\eqref{eq:sigma}).  We also compute $\sigma_{\alpha\beta}$ from the closure
$\tau_{ij}$ extended to 4-dimensions using $u^{\alpha}\sigma_{\alpha\beta}=0$. 
For the NS evolution, the agreement between local viscous heating rate
and observed entropy increase is quite good, as it should be.  The approach
of the right-hand side to zero is particularly notable, since the shear
scalar is an invariant measure of shear, and hence its disappearance of
the approach to uniform rotation.

\begin{figure}
\includegraphics[width=\columnwidth]{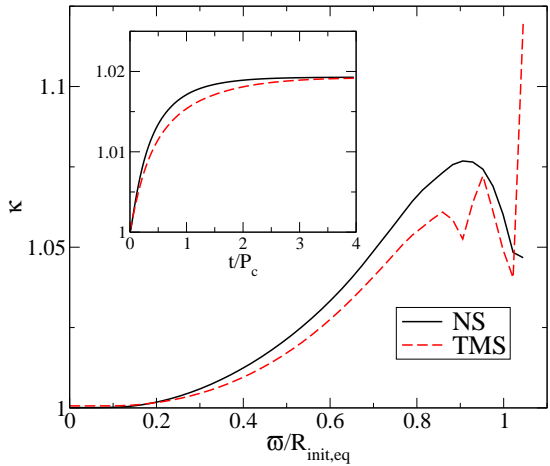}
\caption{Entropy variable $\kappa=P/\rho_0^{\Gamma}$, where $\Gamma=2$,
  for NS and TMS runs.  {\it Main plot:}  The equatorial profile of
  $\kappa$ after 4 initial central rotation periods.  Radii beyond the
  initial equatorial radius have low density and less accurate thermal
  evolution.  {\it Inset:} Evolution of $\kappa$ for a representative
  equatorial tracer particle starting at around 0.6$R_{\rm eq,init}$.}
  \label{fig:heating-cmp}
\end{figure}

For the TMS evolution, we see clearly that the effects of $\tau$
turn off while the covariant $\sigma$ is still nonzero, which we also
noted in discussing the angular velocity evolution.  However, the
results again qualitatively match expectations.  Fluid elements heat
as angular momentum is transported outward.  It should be emphasized
that the disagreement between left and right-hand sides does not in
itself indicate an error in the TMS method; neither of these right-hand
sides is the exact entropy generation rate for the TMS stress; they only
explain why viscous effects turn off while a small amount of invariant
shear remains (because the $\tau$ stress has nearly vanished).  In
particular, it does not indicate that a deficit of heating is causing
energy to disappear.  Viscosity will tend to convert kinetic energy to
heat; TMS in leaving residual shear may simply transfer slightly less energy
without large violations of energy conservation.  For self-gravitating systems,
this analysis is complicated by changes in gravitational potential energy,
and in general relativity total energy is only defined globally by the
asymptotic metric.  Unfortunately, our metric evolution is not accurate enough
to study small changes in total energy that might arise from the TMS terms not
coming from a covariant divergence of a 4D stress tensor.  With our version of
the energy equation, we are at least guaranteed energy conservation in the
Newtonian limit.

To study the viscous heating further, Figure~\ref{fig:heating-cmp} compares
the variable $\kappa = P/\rho_0^{\Gamma}$, a function of entropy initially
equal to one throughout the star, for NS and TMS evolutions.  The entropy at
the final time is higher for the NS case.  This is partly the effect of
greater viscous heating, but when we look at heating at tracer particle locations
(see the figure's inset), we see that the heating of individual fluid elements
is more similar for NS and TMS than the entropy profile would suggest.  The
reason is that the stellar interior expands a bit more in the TMS case than in
the NS case, so that the tracked fluid elements end up at larger radii.  To
speak loosely, it is better to think of the TMS entropy profile as ``shifted
to the right'' compared to the NS profile, rather than as ``shifted down''.
By tracking specific entropy rather than internal energy, we eliminate the
effects of work energy expanding or compressing the fluid.

\subsection{Late-time three-dimensional evolution}
We next evolve the star on a 3D grid to determine non-axisymmetric 
stability. 
In addition to the TMS and NS evolutions, 
we evolve the TMS with $k_1=k_2=0$ for equations~\eqref{eq:turb-visc-dtau}
and~\eqref{eq:turb-visc-dS}. 
We let $\eta=0.05P$, and run the evolution to the viscous timescale to reach 
an equilibrium of constant angular momentum and heating.

Heating is measured by entropy generated, with entropy defined as
\beq
S=\rho_*\log\bigg(\frac{P}{\rho_0^2}\bigg)H\big(\rho-2\times10^{-6}\kappa^{-1}c^2\big)
\eeq
with $H$ being the Heaviside function, to cutoff unphysical entropy growth
in the low density atmosphere.

\begin{figure}
\includegraphics[width=\columnwidth]{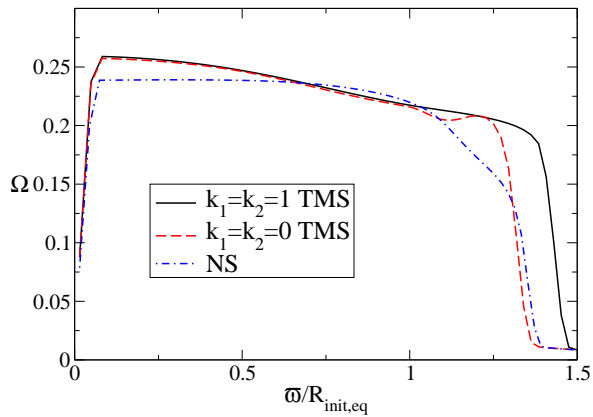}
\caption{Azimuthally averaged angular velocity along equitorial radius at
  time $t=6.6P_c$.}
  \label{fig:3DOmegaAvg}
\end{figure}

\begin{figure}
\includegraphics[width=\columnwidth]{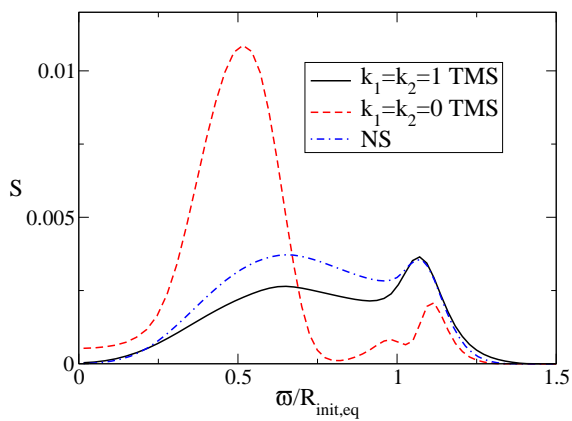}
  \caption{Azimuthally averaged entropy along equatorial radius at time $t=6.6P_c$.}
  \label{fig:3DEntropyAvg}
\end{figure}

\begin{figure*}
  \includegraphics{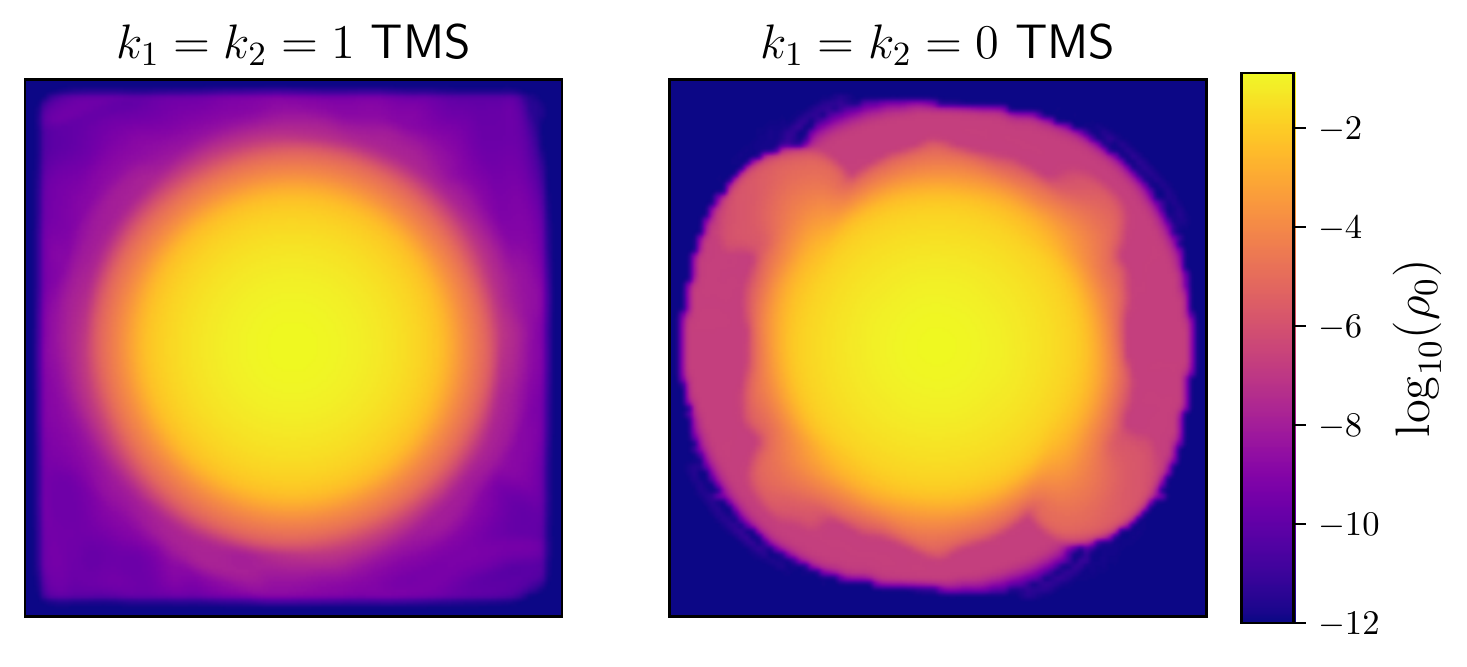}
  \caption{Snapshots of equatorial density at time $t=10.4P_c$ for TMS simulations with $k_1=k_2=1$ and $k_1=k_2=0$ in equations~\eqref{eq:turb-visc-dtau}
    and~\eqref{eq:turb-visc-dS}.  The growth of the $L=4$ mode for $k_1=k_2=0$
    is visible.}
  \label{fig:3DDensity}
\end{figure*}

In figure ~\ref{fig:3DOmegaAvg}, we can see similar behavior in the angular
velocity as the axisymmetric
evolution, with the TMS evolutions settling to a non-zero gradient.
The $k_1=k_2=1$ TMS does show additional expansion of low density material,
which is visible in the density comparison in figure~\ref{fig:3DDensity}.

The entropy in figure \ref{fig:3DEntropyAvg}
shows a very similar behavior between the $k_1=k_2=1$ TMS and NS evolutions, as
expected. In contrast, there is large difference in location and magnitude,
of heating between
the $k_1=k_2=0$ TMS and $k_1=k_2=1$ TMS evolutions, with additional localized heating closer to the core
for the $k_1=k_2=0$ TMS evolution.
The entropy peak of the $k_1=k_2=0$ does drift slowly outward as the simulation progesses.

When we implement the $k_1=k_2=0$ TMS method in SpEC, we observe a $L=4$ mode instability
that results in unphysically strong winds and outflows of low density material.
When the matter reaches the boundary, we are forced to terminate the simulation.
With a domain size of 3.6$R_{\rm eq}$, we did convergence testing with 59, 74, 115, and 144
grid points.
Increasing domain resolution does result in a delayed instability appearance
time, but the growth timescale remains relatively constant for all resolutions, and
is always slightly longer than the timescale needed to reach an equilibrium angular
velocity profile in the star.

Recently, Nedora~{\it et al}~\cite{Nedora:2019jhl} have carried out binary neutron star merger
simulations using Radice's TMS transport.  They find that spiral density waves in
the postmerger remnant generate a wind that could cause a blue kilonova like that
observed as the AT2017gfo counterpart to the binary neutron star merger event
GW170817.  Although the instability reported here may have been present in their
simulations, and a comparison run with $k_1=k_2=1$ is recommended, it is very
unlikely that this would significantly alter the conclusions of that paper.
The existence of spiral waves in post-merger remnants is confirmed by prior
purely hydrodynamical simulations~\cite{Shibata00b,Shibata:2006nm,Bernuzzi:2013rza,bernuzzi:16,Kastaun:2014fna,Paschalidis:2015mla,East:2015vix,Lehner:2016wjg,radice:16onearm}.  Also, Nedora~{\it et al} find that
TMS transport only enhances the outflow mass by $\sim$25\%.

\section{Tests on a Black Hole Accretion Torus System}
\label{sec:disktest}

The astrophysical system most commonly modeled using a phenomenological viscosity
is, of course, disk accretion onto a star or compact object.  For accretion to occur, angular momentum must be transported outward.  It is,
thus, important to study the behavior of different momentum transport treatments
in an accretion disk system and note any major differences.

We evolve a Fishbone-Moncrief torus~\cite{Fishbone:1976}, for which $u^tu_{\phi}$
(roughly, the specific angular momentum) is constant, in our case set to 4.1$M$, where
$M$ is the mass of the black hole.  At
the center is a black hole with dimensionless spin 0.9.  The disk mass is assumed
to be much smaller than that of the black hole, so the spacetime is set to the
Kerr solution in Kerr-Schild coordinates and not evolved.  The disk inner and outer
initial radii are 4.5$M$ and 36$M$, respectively.  The
initial density maximum is at a ring of radius 10$M$.  The
gas of the disk is modeled with $\Gamma=4/3$ equation of state.  The disk is
initially isentropic and obeys a $\Gamma=4/3$ polytropic law.  (Once evolution
begins, the gas will heat.)  All disk mass output below is scaled to the disk's initial
baryonic mass, which can therefore be taken to be one.  This particular system
is not designed to closely model any particular astrophysical scenario, although
a high compaction of the disk (as measured by radius of maximum density divided
by black hole mass) is chosen to be similar to tori encountered in binary post-merger
simulations.

We evolve this system using both TMS (with $k_1=k_2=1$) and NS momentum transport using an alpha viscosity
with $\alpha_{\rm visc}=0.03$.  For TMS simulations, this corresponds to a mixing length
\begin{equation}
  \label{alpha_mix}
  \ell = \alpha_{\rm visc} c_s/\Omega_K\ ,
\end{equation}  
with $\Omega_K$ the Keplerian angular velocity.  Viscosity
is suppressed by an exponential factor for density less than $10^{-4}$ of the initial
maximum, and for gas at radii less than $3M$.  We evolve on
a 2D polar grid with 300 radial points and 256 angular points, uniformly spaced in
the standard accretion grid variables
\begin{align}
 r &= \sqrt{x^2 + z^2} = e^{x_1}, \\
 \theta &= \pi x_2 + \frac{1}{2}(1-h)\sin(2\pi x_2)\ ,
\end{align}
where for these simulations we use $h=0.5$.  The grid covers the range $1.32M\le r\le 2000M$,
$0<\theta<\pi$.  A lower-resolution of $200\times 168$ gives similar results.

We evolve for 100,000$M$.  This is long enough for
the baryonic mass on the grid to drop to 20\% of its initial value for the NS run, 10\%
for the TMS run.  We terminate at this
time because by this time the outer boundary
is no longer sufficiently far.  (This can be seen from the outflow.  At earlier times,
flow through the outer boundary is entirely unbound.  Late in the evolution, the outgoing
mass flux has unbound and weakly bound components, and by $t\approx 10^5M$, the latter has become
comparable to the former.)

\begin{figure}
\includegraphics[width=\columnwidth]{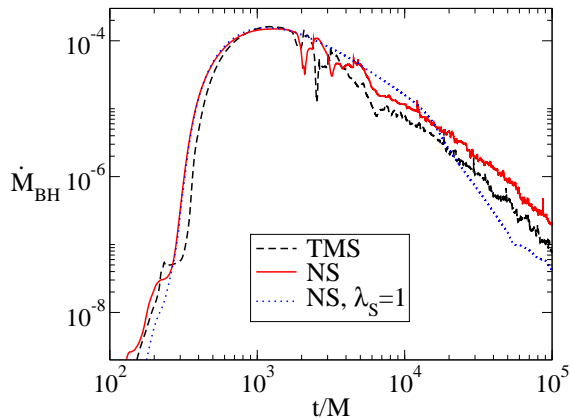}
  \caption{The accretion rate, defined as the fraction of the total baryonic mass of the
    disk accreted into the black hole per interval $M$ of time, for TMS, NS, and NS with
    turbulent heat flux. }
  \label{fig:AccretionRate}
\end{figure}

\begin{figure}
\includegraphics[width=\columnwidth]{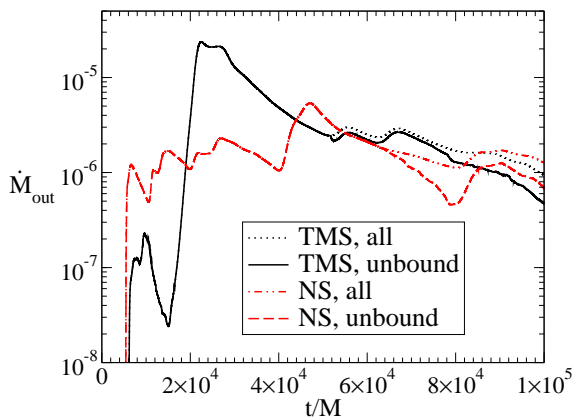}
  \caption{The outflow rate, defined as the fraction of the total baryonic mass of the
    disk leaving the outer boundary per interval $M$ of time, for TMS and NS runs.  We
    plot separately the total outflow of mass and the outflow of unbound mass.  Unbound
    matter is here defined as $u_t<-1$; defining it as $hu_t<-1$ has an insignificant
    effect on the unbound outgoing flux.}
  \label{fig:OutflowRate}
\end{figure}

The baryonic mass flow rate into the black hole and out of the outer boundaries are plotted in
figures~\ref{fig:AccretionRate} and~\ref{fig:OutflowRate}.  The flow rate into the black hole
is seen to be fairly insensitive to the momentum transport method used.  Over the evolved time,
50\% of the baryonic mass accretes into the black hole in the TMS simulation, 60\% in the
NS simulation.  While the difference is not negligible, few would expect any simple phenomenological
model of subgrid turbulent momentum transport to be more accurate than a few tens of percent.
At late times, the accretion rate falls off roughly like a power law $\dot{M}\propto t^{-n}$ where
$1.7\leq n\leq 2$.

The outflow rates show rather larger differences, mostly because of a single large burst of
unbound ejecta in the TMS simulation that is much smaller in the NS evolution.  Over the
evolved time, 17.6\% of the original baryonic mass is ejected from the outer boundary in the
NS evolution:  15.8\% unbound and 1.8\% bound.  For the TMS evolution, 41.1\% of the original
baryonic mass leaves the outer boundary:  39.4\% unbound and 1.7\% bound.  In fact, the slightly
higher accretion rate into the black hole in the NS case might be mostly due to the larger mass
remaining in the disk that did not suffer this one-time ejection.

As a check on whether these differences exceed numerical errors, we evolved the
NS case at two other resolutions, with 0.7 times and 1.3 times the number of
gridpoints in both radial and angular directions, and we use the differences
between resolutions to estimate truncation error.  Matter inflow into the
black hole shows weak dependence on resolution, with errors of a few percent.
The outflow mass is more difficult to resolve and may have error of almost 30\%.
We also reran TMS at the lower resolution and found a nearly 20\% difference in
outflow mass.  Thus, the uncertainty in outflow measurements is tens of
percent, which, while not ideal, is still significantly smaller than the difference
between NS and TMS.  We have also checked to see if our results are sensitive to
the abruptness with which we begin to apply momentum transport.  We perform an additional
NS run in which the viscous parameter $\alpha_{\rm visc}$ is smoothly turned as
$\alpha_{\rm visc}(t) = 0.03(1-e^{-t/300})$.  This leads to a 2\% increase in accreted
mass and roughly 10\% decrease in outflow mass, so this effect also is dwarfed by the TMS-NS
difference.

Of the two methods TMS involves fewer operations to take a timestep,
but the NS  runs are found to be about a factor of two faster because
of the adaptive timestepping used by the SpEC code, which uses larger
timesteps for NS runs to achieve the same time differencing accuracy.

\begin{figure}
\includegraphics[width=\columnwidth]{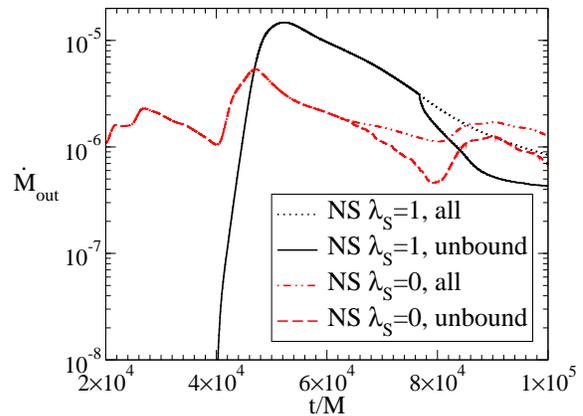}
  \caption{Outflow (total and unbound-only) for NS runs with and without turbulent
    heat conduction.}
  \label{fig:HeatFluxes}
\end{figure}

Figures~\ref{fig:AccretionRate} and~\ref{fig:HeatFluxes} show the
effect of heat fluxes from entropy diffusion, using $\lambda_S=1$.
The disk is initially isentropic, but viscous heating leads to higher
entropy in the interior of the disk.  Heat fluxes therefore supplement
vertical convection and transport energy toward the top and bottom of
the disk.  The main effect of this is to increase the outflow through
the outer boundary to 30.3\% of the initial baryonic mass (28.7\% of
the initial disk mass in unbound outflow).

The production of unbound matter begins differently in simulations
with vs. without heat fluxes.  In both cases, there is very little
unbound material for the first $\sim 2000M$.  Without heat conduction,
viscous heating produces a fairly distinct high-entropy region in the
equator which advects inward.  When it approaches the inner edge of
the disk, it expands vertically, and the first large mass of unbound
matter is ejected from the inner disk region in the polar direction.
With heat conduction, the entropy entropy gradient remains smoother
and shallower ($\kappa\equiv P/\rho_0^{\Gamma}$ is about a factor of
two smaller at the equator than in runs without conduction).  Heat
transport is thus primarily by subgrid-scale convection rather than
large-scale convection.~\footnote{In addition to correcting for the
  loss of small-scale convection due to resolution limits, the
  conduction terms might additionally serve the purpose of correcting
  artifacts of imposing axisymmetry in the evolution, since turbulent
  energy does not cascade to smaller scales in 2D as it does in 3D.}
While a high-entropy region does advect inward at the same time as in
the conduction-free simulations, the perturbation of the inner disk is
far less violent, and only a small mass becomes unbound at this time.
Instead, the massive outflow begins later and starts at the outer
disk, leading to an equatorially concentrated initial burst.  Longer
simulations would be needed to determine if the angular distribution
of the cumulative ejecta is very different depending on whether heat
flux terms are included, but our test suggests that this could be a
possibility in some cases. 

Finally, we perform a demonstration of the possible effect of particle diffusion on the outflow
composition.  We introduce a composition variable $Y$ which is advected by the flow.  Because it
does not enter into the equation of state, it does not affect the evolution (except at the level
of truncation error in the time discretization, if the evolution of $\rho Y$ is allowed to
occasionally control the adaptive timestep).  Ejecta composition is of great interest in
post-merger simulations because of its connection to kilonovae and r-process nucleosynthesis.
Since our simulation lacks neutrino interactions, it should be considered only a demonstration
of another possibly significant influence.

\begin{figure}
\includegraphics[width=\columnwidth]{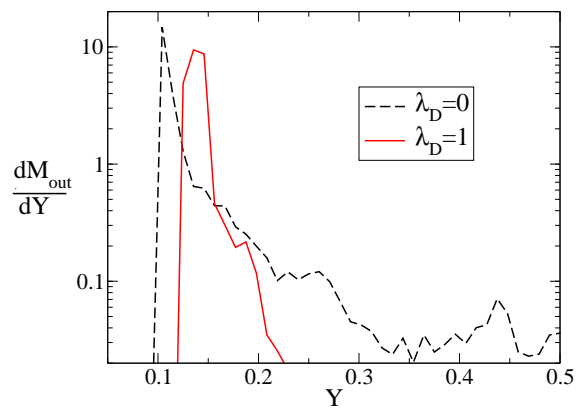}
  \caption{The distribution of composition variable $Y$.  $\frac{dM}{dY}$ 
    is defined such that the baryonic mass $dM_0$ within $dY$ of $Y$ is
    $dM_0 = \frac{dM}{dY}dY$.}
  \label{fig:outflowYe}
\end{figure}

We initialize $Y$ as
\begin{equation}
  Y =
  \begin{cases}
    0.5-6\rho_{\rm 0,init}/\rho_{\rm 0,init,max} ,& \text{if } \rho_{\rm 0,init}/\rho_{\rm 0,init,max} < \frac{1}{15} \\
    0.1,              & \text{otherwise}
    \end{cases}
\end{equation}  
The idea of using a simple analytic form with higher $Y$ at low
densities and $Y=0.1$ at high densities was taken
from~\cite{Fujibayashi:2020qda}, which in term is a rough fit to the
electron fraction in binary neutron star post-merger accretion disks.
We evolve without particle diffusion ($\lambda_D=0$) and with it
($\lambda_D=$1).  Both simulations use NS transport and compute the
mean free path from $\alpha_{\rm visc}$ as in Eq.~\eqref{alpha_mix}.  We
integrate the mass flux passing through the surface $r=800$M in $Y$
bins to get the total mass outflow as a function of $Y$, which is
plotted in Fig.~\ref{fig:outflowYe}.  In this example, the effect of
particle diffusion is to shift the distribution peak to higher $Y$ and
to make it narrower.  One noticeable difference is that the high-$Y$
tail of the outflow distribution disappears when particle diffusion is
added.  Recall that in this simple test, the composition variable does
not affect the equation of state and indeed plays no role in the
hydrodynamics at all, so the same fluid elements are ejected in both
simulations.  However, there is a fairly strong $Y$ gradient in the
initial disk near the low-density layers at the top and bottom edges
of the disk, and this is where the high-$Y$ material is.  The
diffusion term causes the $Y$ composition variable to ``bleed into''
the disk, significantly lowering $Y$ near the surface of the disk. 
This flow of particles will also raise $Y$ in the disk interior, but
the composition flux will lower $Y$ near the surface more than it will
raise $Y$ in the interior, because the density is higher in the latter
region.

Overall, the effect of smoothing the $Y$ distribution is that $Y$ does
not vary as much in the density layers that provide the ejecta.  Of
course, the effect may be different for different composition
distributions or in the presence of composition source terms
(e.g. neutrino interactions).  Interestingly, a detailed comparison of
disks evolved with $\alpha$-viscosity vs magnetohydrodynamics found an
opposite effect, that the MHD run had wider composition distribution
at a lower peak~\cite{Fernandez:2018kax}.  The diffusive effects of
magnetorotational turbulence is certainly one effect present in MHD
simulations but not viscous simulations without particle diffusion,
although in this case outflows driven by large-scale magnetic fields
may have been the more important difference.

\section{Conclusion}
\label{sec:conclusion}


Even for a given choice of the effective viscosity $\eta$ there is some freedom in how one
adds momentum transport to the relativistic Euler equations. In this manuscript, we perform
detailed comparison of two models currently in use in numerical relativity simulations: Shibata
{\it el al}'s NS model, and Radice's TMS model. We also propose an improvement to the TMS
model: the addition of physically motivated terms that guarantee that the stress-energy
tensor remains a spatial tensor in the fluid rest frame, and prevents slowly-growing instability
to appear in some test problems. The main objective of these models in merger simulations
has been to provide angular momentum transport in the post-merger remnant. We find
that the NS, original TMS, and modified TMS model fortunately behave very similarly in that respect,
at least within the expected uncertainties of a mean field turbulence model. However, we find
significant differences in viscous heating between the original TMS and the other two models, while
all models provide different results for the momentum transport-driven ejecta
mass in disk simulations. It is already known that disk outflow
masses depend on $\alpha_{\rm visc}$.  To this, we add that even if a ``correct'' $\alpha_{\rm visc}$
were known, outflows would still depend on the transport formalism. 

As the TMS formalism is not 4-dimensionally covariant, its results might not apply for arbitrary
foliations of the spacetime.  We doubt, however, that the slicing choices usually used
by numerical relativists would lead to dramatically different foliations, or that the gauge-dependence of
TMS would impact numerical results more than the approximations inherent to any mean field model.
Additionally, as the TMS model is simpler to implement and computationally less expensive than the 
NS model (at least on a per timestep basis), it certainly remains very useful to numerical
simulations. The improvements  to the TMS model proposed in this manuscript add new terms
to the evolution equations, but without increasing the complexity of the evolution algorithm itself,
or meaningfully impacting the cost of simulations. They should thus be reasonably simple to implement
in any TMS-based code.

Mean-field models of subgrid transport effects provide an economical way to
explore deep into the post-merger regime, although of course they cannot
replace a more limited number of expensive high-resolution simulations.  These
models could easily be improved beyond what we have attempted here.  An adequate
model of subgrid effects would have to include heat transport, and it should
also account for the effective pressure from turbulent stresses which has
proved to be potentially quite important in the supernova core collapse
problem~\cite{couch:15a}.  It would also be interesting to add the evolution of the
large-scale magnetic field--even if the magnetorotational instability is subgrid scale--
in order to incorporate large-scale magnetohydrodynamic effects such as magnetic braking
and jet collimation.  In the presence of subgrid turbulence, the induction
equation for the mean magnetic field would itself need to be suitably augmented to
include subgrid electromotive force terms, as is done in dynamo
modeling~\cite{Brandenburg:2004jv}
and even in a few relativistic simulations~\cite{Giacomazzo:2014qba,Sadowski:2014awa}.
If the mean field grows large enough
to resolve the magnetorotational instability, then one would more correctly be in
a regime for large eddy rather than mean field modeling.

One might question the point of improving models which at best capture their effects
to no better than order of magnitude anyway.  It is useful for at a couple of reasons.
First, one is able to establish the sensitivity of particular outputs to various transport
effects, as we have done with disk outflow composition, so that it is known what effects
are most important for high-resolution simulations to capture.  Second, these simplified
models play an important role in interpreting high resolution results, guiding the
inevitable tradeoff between exactness and human intelligibility.

\begin{acknowledgements}
We are thankful to David Radice for many discussions on the TMS formalism,
and advice on its numerical implementation in our code.
M.D gratefully acknowledges
support from the NSF through grant PHY-1806207.
The UNH authors gratefully acknowledge
support from the DOE through Early Career award de-sc0020435,
from the NSF through grant PHY-1806278, and
from NASA through grant 80NSSC18K0565.
J.J. gratefully acknowledges
support from the Washington NASA Space Grant Consortium, NASA Grant NNX15AJ98H.
L.K. acknowledges support from NSF grant
PHY-1606654 and PHY-1912081. F.H. and M.S. acknowledge support from NSF Grants
PHY-1708212 and PHY-1708213. F.H., L.K. and M.S. also thank
the Sherman Fairchild Foundation for their support.
\end{acknowledgements}

\bibliography{References/References.bib}

\end{document}